\documentclass[aps,prl,twocolumn,superscriptaddress,showpacs]{revtex4}

\usepackage{graphicx}
\usepackage{epstopdf}
\usepackage{amssymb}
\usepackage{amsmath}

\begin{document}

\title{Observation of Two-Neutrino Double-Beta Decay in $^{136}$Xe with EXO-200}



\author{N.~Ackerman}
\altaffiliation{Now at Dept. of Radiation Oncology, Stanford University, Stanford CA, USA}
\affiliation{SLAC National Accelerator Laboratory, Stanford CA, USA}
\author{B.~Aharmim}
\affiliation{Physics Department, Laurentian University, Sudbury ON, Canada}
\author{M.~Auger}
\affiliation{LHEP, Albert Einstein Center, University of Bern, Bern, Switzerland}
\author{D.J.~Auty}
\affiliation{Department of Physics and Astronomy, University of Alabama, Tuscaloosa AL, USA}
\author{P.S.~Barbeau}
\altaffiliation{Corresponding author: psbarbea@stanford.edu}
\affiliation{Physics Department, Stanford University, Stanford CA, USA}
\author{K.~Barry}
\author{L.~Bartoszek}
\affiliation{Physics Department, Stanford University, Stanford CA, USA}
\author{E.~Beauchamp}
\affiliation{Physics Department, Laurentian University, Sudbury ON, Canada}
\author{V.~Belov}
\affiliation{Institute for Theoretical and Experimental Physics, Moscow, Russia}
\author{C.~Benitez-Medina}
\affiliation{Physics Department, Colorado State University, Fort Collins CO, USA}
% ...

\collaboration{EXO Collaboration}

\date{17 November 2011}

\begin{abstract}

We report the observation of two-neutrino double-beta decay in $^{136}$Xe
with $T_{1/2}$=2.11$\pm$0.04(stat.)$\pm$0.21(sys.)$\times$10$^{21}$~yr.  This second order process, predicted 
by the Standard Model, has been observed for several nuclei but not for
$^{136}$Xe.  The observed decay rate provides new input to matrix element 
calculations and to the search for the more interesting neutrino-less double-beta decay, 
the most sensitive probe for the existence of Majorana particles and the 
measurement of the neutrino mass scale.

\end{abstract}


\maketitle

Several even-even nuclei are stable against ordinary $\beta$ decay but are unstable for
$\beta\beta$ decay in which two neutrons are changed into two protons simultaneously.
As is well known, $\beta\beta$ decay can proceed through several modes. 
The allowed process, the two-neutrino  mode ($2\nu\beta\beta$),
is completely described by known physics; its rate was first evaluated
in~\cite{GoppertMayer}.  Of the other, hypothetical, modes, the neutrino-less decay
($0\nu\beta\beta$) is forbidden in the Standard Model since it violates conservation
of the total lepton number.  Its observation would constitute proof that neutrinos
are Majorana leptons~\cite{Majorana}, unlike all charged fermions that are of
the Dirac type~\cite{BjorkenDrell}.
Moreover, the $0\nu\beta\beta$ decay can proceed only if neutrinos have mass \cite{Schechter}. Consequently, there is an intense  worldwide
program of experiments aiming at observing the $0\nu\beta\beta$ mode.
The relation between the 
$0\nu\beta\beta$ half-life and the average Majorana neutrino mass requires
the evaluation of nuclear matrix elements that, while different from those of
the $2\nu\beta\beta$ decay mode, would benefit from their knowledge.  Indeed, it has 
been suggested~\cite{Petr_et_al} that the theoretical parameters contributing to the 
largest uncertainties in the $0\nu\beta\beta$ matrix element calculation can be derived
from the $2\nu\beta\beta$ decay matrix elements, known once the half-life has been
experimentally measured.   
The half-life of the $2\nu\beta\beta$ decay depends
on details of the nuclear structure that are only known approximately~\cite{Avignone}.  The 
$2\nu\beta\beta$ decay has been observed in all important candidate nuclei \cite{RPP}
with one notable exception, $^{136}$Xe, which until now had only lower limits on 
the half-life~\cite{limits}.  The most stringent published limit would imply a nuclear matrix element noticeably 
smaller than those found for other isotopes.   
From an experimental perspective, the $0\nu\beta\beta$ and $2\nu\beta\beta$ modes
can be distinguished from the study of the energy spectrum of the electrons.  The sum energy spectrum is a resolution-limited line at the Q-value for $0\nu\beta\beta$ (2458~keV
for $^{136}$Xe~\cite{Redshaw}) and
a broad continuum for $2\nu\beta\beta$, so that good energy resolution, along with
the knowledge of the $2\nu\beta\beta$ rate are essential in the search for $0\nu\beta\beta$.

\begin{figure}
\includegraphics[width=3in]{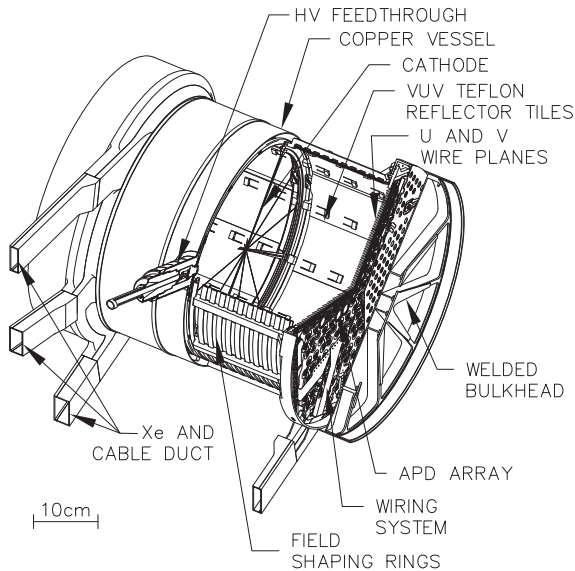}
\caption{Drawing of the EXO-200 TPC.  The chamber contains $\sim$175~kg of liquid Xenon
enriched to 80.6\% in the isotope 136.}
\label{fig:TPC}
\end{figure}

The EXO-200 detector, shown in Figure~\ref{fig:TPC}, is a time projection 
chamber~\cite{TPC} (TPC) using liquid Xe (LXe) both as the source of nuclear 
decays and the detection medium.  
The TPC has the geometry of a cylinder of 40~cm diameter and 44~cm length, with the 
drift field obtained by biasing a cathode grid dividing the cylinder into two 
identical regions.  Each end of the cylinder is flared into a conical section, 
containing two wire grids and one array of $\sim 250$ large-area 
avalanche photodiodes~\cite{APDs} (LAAPDs) that allow for simultaneous readout of 
ionization and scintillation in the LXe.  Wire grids cross at a 60$^{\circ}$ angle, 
providing 2-dimensional localization and energy readout of each charge deposition.  
The third (longitudinal) coordinate is obtained from the time interval between the 
scintillation signal in the LAAPDs and the collection of the charge at the grids.
A set of field shaping rings, lined with reflective teflon tiles, grades the field 
and limits the drift region to two cylinders, each of 18.3~cm radius and 
19.2~cm length.  For the data presented here the cathode bias was set to -8.0~kV, 
providing a field of 376~V/cm, designed to be uniform to within 1\% over the entire fiducial volume.  This low value of the electric field provides more stable operation at the expense of the ionization energy resolution that is not essential for the measurement
of the $2\nu\beta\beta$ mode.

All components used for the construction of the detector were carefully selected 
for low radioactive content~\cite{activity} and compatibility with electron 
drift in LXe. The TPC is mounted in the center of a low-background cryostat filled 
with $\sim 2400$~l of high-purity HFE7000 fluid~\cite{HFE} serving the purpose of 
innermost radiation shield and heat transfer fluid.  At least 50~cm of HFE7000
(with a density of 1.8~g/cm$^3$ at 167~K) separate the TPC from other components. 
The LXe (and the HFE7000 fluid) is held at 147~kPa (1100~torr) and 167~K, with 
possible temperature variations $<\pm 0.1$~K, by cooling the inner vessel of 
the cryostat with a closed circuit refrigerator.  The cryostat is vacuum insulated 
and has a total radial thickness of 5~cm of low background copper.  It is further 
encased in a 25~cm thick low-activity lead shield.  Signals from wire triplets,
spaced 9~mm from each other, and LAAPDs are brought out of the cryostat 
and lead shield, where they are amplified, shaped and digitized at 1~MS/s by 
room temperature electronics. 
The detector infrastructure includes a gas phase recirculation system consisting
of a boiler, a pump, a hot Zr purifier, gas purity monitors~\cite{GPM} and a condenser.
A substantial control system maintains a very small ($<85$~torr) pressure difference 
across the TPC vessel that is built out of 1.37~mm 
thin copper to keep backgrounds low.  A calibration system allows the insertion of 
miniaturized radioactive sources
to various positions immediately outside of the TPC.  

The clean room module housing the cryostat and the TPC is surrounded on four 
sides by an array of 50~mm thick plastic scintillator panels~\cite{KARMEN}.
The array detects muons traversing the lead shielding with an efficiency of 95.9\%.  
EXO-200 is located at a depth of about 1600~m.w.e. in a salt deposit of the Waste 
Isolation Pilot Plant (WIPP), near Carlsbad, NM.  The muon flux at this site was measured~\cite{muon} to be $3.1\times 10^{-7}$~s$^{-1}$cm$^{-2}$sr$^{-1}$.  
A paper describing the EXO-200 detector in detail is in preparation.

For the data presented here EXO-200 was filled with $\sim$175~kg of xenon
enriched to $80.6\pm 0.1$\% in the isotope 136 ($^{\rm enr}$Xe).  The remaining 
fraction (19.4\%) is the isotope 134, the rest of the natural Xe isotopes 
represent negligible contributions.
$^{85}$Kr is a radioactive fission product with $Q_{\beta}=687.1$~keV and 
$T_{1/2}=10.8$~yr that is present in the atmosphere since the nuclear age and 
generally contaminates Xe, as a trace component of natural Kr.  The EXO-200 enriched 
Xe was measured~\cite{Kr_det,Kr_meas} to contain $(25\pm 3) \times 10^{-12}$~g/g 
of natural Kr, substantially less than the typical concentration of 
$10^{-8}$ to $10^{-7}$ found in Xe after distillation from air.

\begin{figure}
\includegraphics[width=3in]{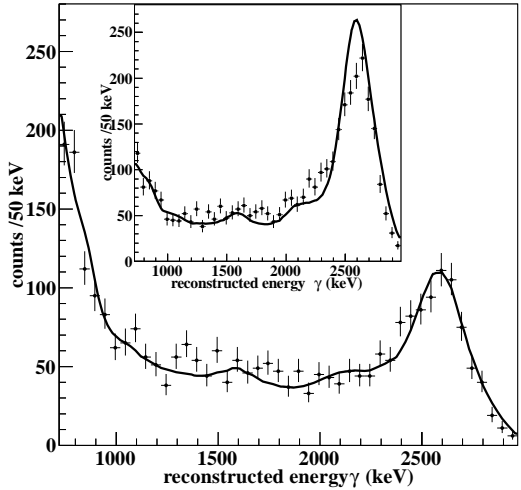}
\caption{Energy spectrum for a $^{228}$Th calibration source at the
mid-plane of the TPC, 3~cm outside the LXe volume.  The intensity (vertical scale) is not fit: the agreement between the Monte Carlo calculations (solid line) and the data tests the accuracy of the simulation against the absolute, National Institute of Standards and Technology traceable, source activity.  The energy scale (horizontal axis) has been corrected for $\tau_e$ at the time of collection and has had separate energy calibrations applied for single- (main plot) and multiple-cluster (inset) interactions. }
\label{fig:agreement}
\end{figure}

The data were collected between May 21, 2011 and July 9, 2011, for a total of  
752.66~hrs of low background running.  During the same period, about two hours 
of every day were devoted to detector calibration using $^{60}$Co and $^{228}$Th
sources.  A specific source was inserted each day at one out of five ``standard'' 
positions near the TPC.  Typically a source was used to scan these positions 
over a week and then replaced with a different one.  
Data analysis is performed by two independent groups, providing cross checks
of the results.
The detector calibration procedure begins by fitting the energy spectra from 
the sources to obtain an electron lifetime ($\tau_e$) in the LXe and an overall 
correspondence between the charge and the energy deposited in the detector.  After an initial phase of recirculation of the Xe $\tau_e$ reached 250~$\mu$s and
it remained between 210 and 280~$\mu$s in the data set used here (the maximum drift time
at the field used here is $\sim 100$~$\mu$s).  $\tau_e$ values are obtained by minimizing the energy 
resolution for source calibration events, occurring at locations randomly distributed over the entire LXe volume.  The dispersion in the $\tau_e$ measurements is incorporated in the 
systematic uncertainty.  The daily calibration schedule makes it possible to track and correct 
for changes in $\tau_e$.  

For this initial analysis only the ionization signal is used to measure the energy.  The scintillation signals recorded by the LAAPDs are 
used to establish the time of the event, identify $\alpha$ particles 
by their higher light-to-charge ratio compared to electron-like events
and measure $\alpha$ energies.  
The combined use of scintillation and ionization, to obtain the best energy 
resolution, is under development.  

The ability of the TPC to reconstruct energy depositions in space is used to 
remove interactions at the detector edges where the background is higher.
 It also discriminates between single-cluster depositions, characteristic of 
$\beta\beta$ and single $\beta$ decays in the bulk of the Xe, from multi-cluster ones,
generally due to $\gamma$-rays that constitute the majority of the background.  In the present analysis, such discrimination only employs one spatial coordinate ($\approx$15~mm separation) and the time coordinate ($\approx$17~mm separation).  
The fiducial volume used here contains 63~kg of $^{\rm enr}$Xe ($2.26\times10^{26}$ 
$^{136}$Xe atoms).  The detector simulation, based on GEANT4~\cite{GEANT4},
reproduces the energy spectra taken with calibration sources well.  This also applies to the single- to multi-cluster assignment obtained with the
external calibration sources, as illustrated in Figure~\ref{fig:agreement} for
the case of $^{228}$Th.

\begin{figure}
\includegraphics[width=3in]{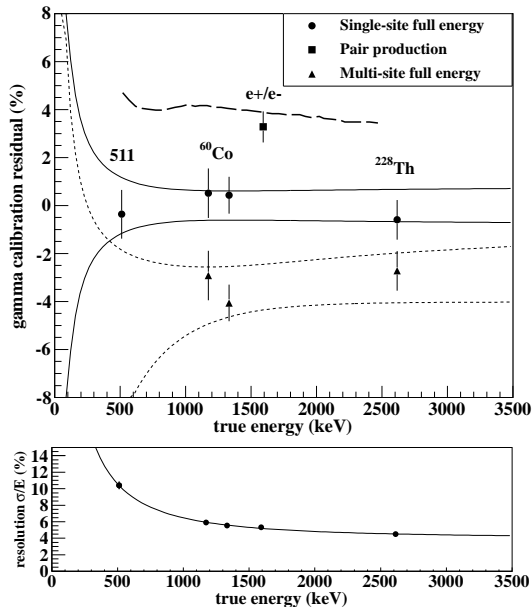}
\caption{Top: fractional residuals between the energy calibration points and the linear model 
discussed in the text.  The single- (solid line) and multi-cluster (dotted line) uncertainty bands are systematic, stemming from the finite
accuracy of the position reconstruction and the $\tau_e$ correction.  The thick dashed 
line represents the central value of the shift predicted by the simulation for 
point-like energy depositions.  Bottom: measured energy resolution (points) along 
with a parameterization (line).}
\label{fig:Residuals}
\end{figure}

Four full absorption $\gamma$ calibration peaks, spanning the energy region of 
interest for this analysis, are derived from the $^{60}$Co and $^{228}$Th sources: 
1173~keV, 1332~keV, 2615~keV and 511~keV (annihilation
radiation).  The three high energy $\gamma$s provide both single-cluster and 
multiple-cluster event samples.  The energy scale is found to be slightly 
($\sim$~4\%)
different in the two samples because of the non-zero charge collection threshold 
on individual wire triplets.  An additional calibration energy at 1592~keV is provided by selecting ionization 
sites produced by e$^+$e$^-$ pairs from the highest energy $\gamma$s. The nature 
of these energy depositions, however, is different from the others, being produced 
directly by ionization in a smaller volume.  This type of deposition is analogous
to that expected from $\beta\beta$ decay and is found to be slightly different from
single-cluster depositions from $\gamma$s.  This shift is well reproduced by
the simulation, once induction between neighboring wire triplets and other electronics
effects are taken into account.  After correcting for these two shifts and the (slowly) time-varying $\tau_e$ 
the energy scale fits well to a linear function.  The fractional residuals from this process are shown in 
the top panel of Figure~\ref{fig:Residuals}.  

The measured energy resolution is $\sigma_E = 4.5$\% at 2615 keV.  
A parameterization of the resolution function is incorporated into the simulation, as 
shown in the bottom panel of Figure~\ref{fig:Residuals}.  
The analysis uses an energy threshold of 720~keV, chosen so that both the trigger and event reconstruction are fully efficient.
Probability Distribution Functions (PDFs) for each source and position
are generated by means of Monte Carlo simulation and compared to the single- and multiple-cluster data (see Figure \ref{fig:agreement}).  This 
procedure reproduces the activities of the external calibration $\gamma$ 
sources to within $\pm$8\% of their known 
activities.  

\begin{figure}
\includegraphics[width=3.5in]{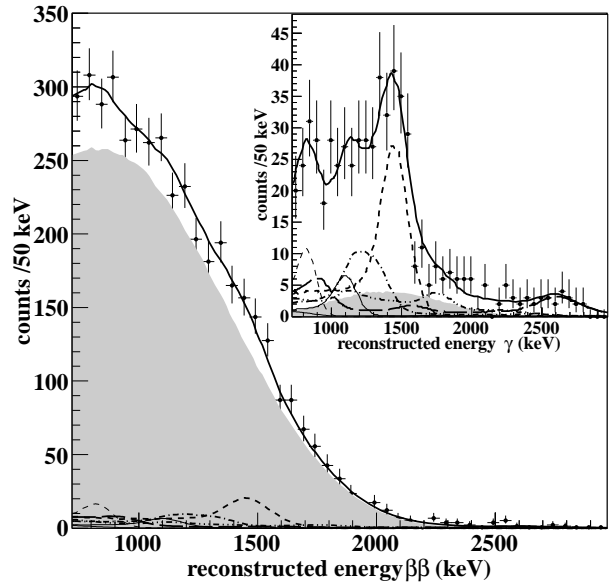}
\caption{Energy distributions from 752.66~hrs of EXO-200 single-cluster events (main panel) 
and multi-cluster events (inset).  
The result of a likelihood fit to a model including the $2\nu\beta\beta$ decay and 
several backgrounds is shown (solid line) along with the $2\nu\beta\beta$ component 
(shaded region) and some prominent background components at the radius of the TPC vessel
($^{232}$Th, long dash; 
$^{40}$K, dash; $^{60}$Co, dash-dot; $^{54}$Mn, thin dash; $^{65}$Zn, thin solid;
$^{238}$U chain in equilibrium, dash double-dot).  Other background components fitting to 
negligible amounts are not shown, for clarity.  The energy scale used for the main 
panel is consistent with that of single-cluster, $\beta$-like events while the scale 
of the inset is consistent with the multi-cluster events it represents.  The combined
$\chi^2$/ degrees of freedom between the model and the data for the two binned 
distributions shown here is 85/90.}
\label{fig:SpectralFit}
\end{figure}

The data collected during low background running requires
only two selection cuts to remove modest backgrounds.  Cosmic-ray induced 
backgrounds are rejected by removing events preceded by a veto counter hit 
in a 5~ms window.  This cut removes 124 events introducing a dead time 
of 0.12\%. 
The decay rate of $^{222}$Rn is independently determined to 
be 4.5$\pm$0.5 $\mu$Bq~kg$^{-1}$ from an $\alpha$-spectroscopy analysis performed using only scintillation signals, consistent with $\beta$--$\alpha$ and $\alpha$--$\alpha$ time 
coincidence analyses.  Similarly, $^{220}$Rn is constrained to $<0.04 \; \mu$Bq~kg$^{-1}$ (90\%~CL).  In the data set 72 $\beta$--$\alpha$ coincidences are removed.  The implementation of this cut introduces a 6.3\% dead time due to spurious ionization or scintillation signals.  
Events are then classified as single-
or multi-cluster and energy spectra are obtained for these two classes, as
shown in Figure~\ref{fig:SpectralFit}.  The spectra are simultaneously 
fit to PDFs for the $2\nu\beta\beta$ decay signal (65\% of which is above threshold) and various 
backgrounds using an un-binned maximum likelihood method.  The $2\nu\beta\beta$ PDF is produced using the Fermi function calculation given in~\cite{VogelFermiFunction}.  The detector simulation predicts a small fraction of the $2\nu\beta\beta$ decay signal to be classified as the multi-cluster type because of brehmmstrahlung as well as charge collection effects.  Background models are 
developed for various components of the detector, inspired by screening
of materials performed at the time of the detector's construction and 
by estimated cosmogenic activation.
As Figure~\ref{fig:SpectralFit} illustrates, the backgrounds involving $\gamma$
rays are readily identified by their clear multi-cluster signature, while the 
single-cluster spectrum is dominated by a large structure with a shape consistent
with the $2\nu\beta\beta$ decay of $^{136}$Xe.  
The simultaneous likelihood fit to the single and multi-cluster spectra reports a strong signal from the $2\nu\beta\beta$ decay 
(3886 events) and a dominant contamination from $^{40}$K at the location of 
the TPC vessel (385 events).  Other contributions account for a total of less 
than 650 events, each with a very low significance in the fit.
These levels of contamination are consistent with the material screening 
measurements~\cite{activity}.  Taking only the single-cluster events into account the single-to-background ratio is 9.4 to 1.

The $\alpha$-spectroscopy analysis is used to bound any $^{238}$U contamination in the
bulk LXe.  This is important because $^{238}$U decays are followed (with an average delay of 
$\sim 35$~d) by $^{234m}$Pa decays, producing $\beta$s with a 
Q-value of 2195~keV.  The $\alpha$ scintillation spectrum is calibrated
using the lines observed from the $^{222}$Rn chain, obtaining a limit
for $^{238}$U (and $^{234m}$Pa) of $<10$~counts for the data set shown in Figure~\ref{fig:SpectralFit}.    In addition, a study of the production of
fast neutrons resulting in recoils and captures in the LXe as well as thermal 
neutrons resulting in captures is used to bound these backgrounds to $<10$ events
for the data set in the Figure.

The measured half-life of the $2\nu\beta\beta$ decay in $^{136}$Xe obtained
by the likelihood fit is 
$T_{1/2}$=2.11$\pm$0.04(stat)$\pm$0.21(sys)$\times$10$^{21}$~yr,
where the systematic uncertainty includes contributions from the 
energy calibration (1.8\%), multiplicity assignment (3.0\%), fiducial volume (9.3\%)
and $\gamma$ background models (0.6\%), added in quadrature.   
The uncertainty from the energy calibration is estimated using a Monte Carlo 
method scanning calibration constants within the range illustrated in 
Figure~\ref{fig:Residuals} and re-fitting the spectra, weighting the fit 
results by their likelihood value.  The same method is used to quantify 
the effect of the multiplicity assignment.  The fiducial volume uncertainty 
is determined from the fidelity with which calibration events are reconstructed 
within a chosen volume as compared to simulation.  The $\gamma$ background 
model uncertainty is derived from the results of likelihood fits performed 
with a variety of different background hypotheses. 

\begin{figure}
\includegraphics[width=3in]{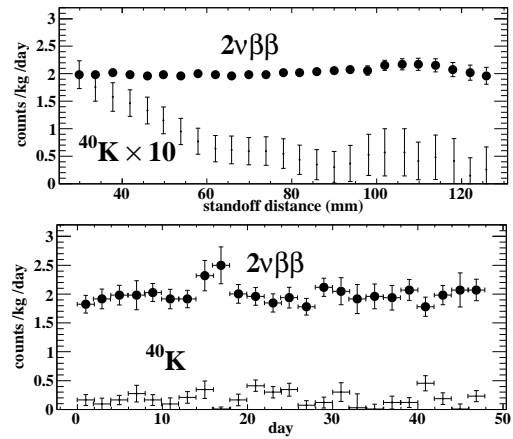}
\caption{Top: measured $2\nu\beta\beta$ decay rate of $^{136}$Xe (large points)
and largest background contribution ($^{40}$K, small points) as a function of 
the standoff distance from detector components.  
Bottom: event rates of $2\nu\beta\beta$ and $^{40}$K decays as a function of time. }
\label{fig:constancy}
\end{figure}

In Figure~\ref{fig:constancy} the fitted values of the $2\nu\beta\beta$ and 
the $^{40}$K background are shown as functions of the event standoff distance
from materials other than the LXe (top panel) and time in the run.  While the 
$^{40}$K is attenuated by the LXe as expected, the  $2\nu\beta\beta$ signal appears to be 
uniformly distributed in the detector and constant in time.

An exhaustive search for $\beta$ emitters with no $\gamma$s, $T_{1/2}>2$~days 
and energies of interest yields only two candidates: $^{90}$Y (supported by 
$^{90}$Sr) and $^{188}$Re (supported by $^{188}$W).   It appears \emph{a priori} 
unlikely that the bulk of the LXe is uniformly contaminated with these isotopes while
simultaneously not showing significant evidence for more common metallic contaminants 
such as those from the $^{238}$U decay chain.      Nevertheless, additional 
test fits are performed by incorporating each isotope separately.  At 90\% C.L. the $2\nu\beta\beta$ rate is reduced by less than 7\% (30\%) for the inclusion 
of $^{90}$Y ($^{188}$Re).  

In conclusion the initial data taking of EXO-200 has provided a clear detection
of the $2\nu\beta\beta$ decay in $^{136}$Xe.  The measured $T_{1/2}$ is significantly
lower than the lower limits quoted in~\cite{limits} and translates to a nuclear
matrix element of 0.019~MeV$^{-1}$, the smallest measured among the $2\nu\beta\beta$ emitters.

\begin{acknowledgments}
EXO-200 is supported by DoE and NSF in the United States, NSERC in Canada, SNF in 
Switzerland and RFBR in Russia.  The collaboration gratefully acknowledges the 
hospitality of WIPP.
\end{acknowledgments}


\end{document}